\documentclass[prd,superscriptaddress,a4paper,showpacs,showkeys,10pt,nofootinbib]{revtex4}
\usepackage{graphicx}
\usepackage{graphics}
\usepackage{epsfig}
\usepackage{dcolumn}
\usepackage{bm}
\usepackage{multirow}
\usepackage{tabularx}
\usepackage{hyperref}
\usepackage{commath}
\usepackage{epstopdf}
\usepackage[T1]{fontenc}
\usepackage{geometry}
\usepackage{color}

\def\be{\begin{equation}}
\def\ee{\end{equation}}

\providecommand{\ee}{e$^+$e$^-$}

\providecommand{\tabularnewline}{\\}

\makeatother

\begin{document}
%
%
\title{Photoproduction of massive gauge bosons in $pp$, $pPb$ and $PbPb$ collisions}



\author{R. O. Coelho}

\email[]{coelho72@gmail.com}

\affiliation{Instituto de F\'{\i}sica e Matem\'atica, Universidade Federal de
Pelotas (UFPel),\\
Caixa Postal 354, CEP 96010-090, Pelotas, RS, Brazil}

\author{V. P. Gon\c calves}

\email[]{barros@ufpel.edu.br}

\affiliation{Instituto de F\'{\i}sica e Matem\'atica, Universidade Federal de
Pelotas (UFPel),\\
Caixa Postal 354, CEP 96010-090, Pelotas, RS, Brazil}



\begin{abstract}
In this letter we present, for the first time, the results for the photoproduction of massive gauge bosons in proton -- proton, proton -- Lead and Lead -- Lead collisions at the  Large Hadron Collider (LHC), High -- Energy LHC (HE -- LHC) and Future Circular Collider (FCC). Predictions for the rapidity distributions and total cross sections are presented. We predict a large number of events in the  rapidity range probed by the LHC detectors, which implies that this process can be used to probe the photoproduction of massive gauge bosons as well to perform the search of Beyond Standard Model physics.
\end{abstract}




\maketitle


The experimental results for photon -- induced processes at Tevatron, RHIC and LHC have  motivated a series of studies that propose  to test  our understanding of different aspects of the Standard Model as well   to constrain possible scenarios of the Beyond Standard Model (BSM) physics using the analysis of different final states produced in  $\gamma \gamma$ and $\gamma h$ interactions at hadronic collisions (For a review see, e.g. Ref. \cite{upc9}). In this letter we will analyze the possibility of improve our understanding about the couplings between the gauge bosons considering the photoproduction of $Z$ and $W$ in $pp$, $pPb$ and $PbPb$ collisions for the energies of the next run of the LHC,  as well of the High -- Energy LHC  \cite{he_lhc} and Future Circular Collider \cite{fcc}. In particular, we will estimate, for the first time, the   total cross sections and expected number of events considering the rapidity range covered by the LHC detectors. Our goal is to verify if the experimental analysis of this process is feasible, which will  allow to use it to  test  the Standard Model predictions, as well to search  effects from new physics.





\begin{figure}
\begin{tabular}{cc}
\hspace{-1cm}
{\psfig{figure=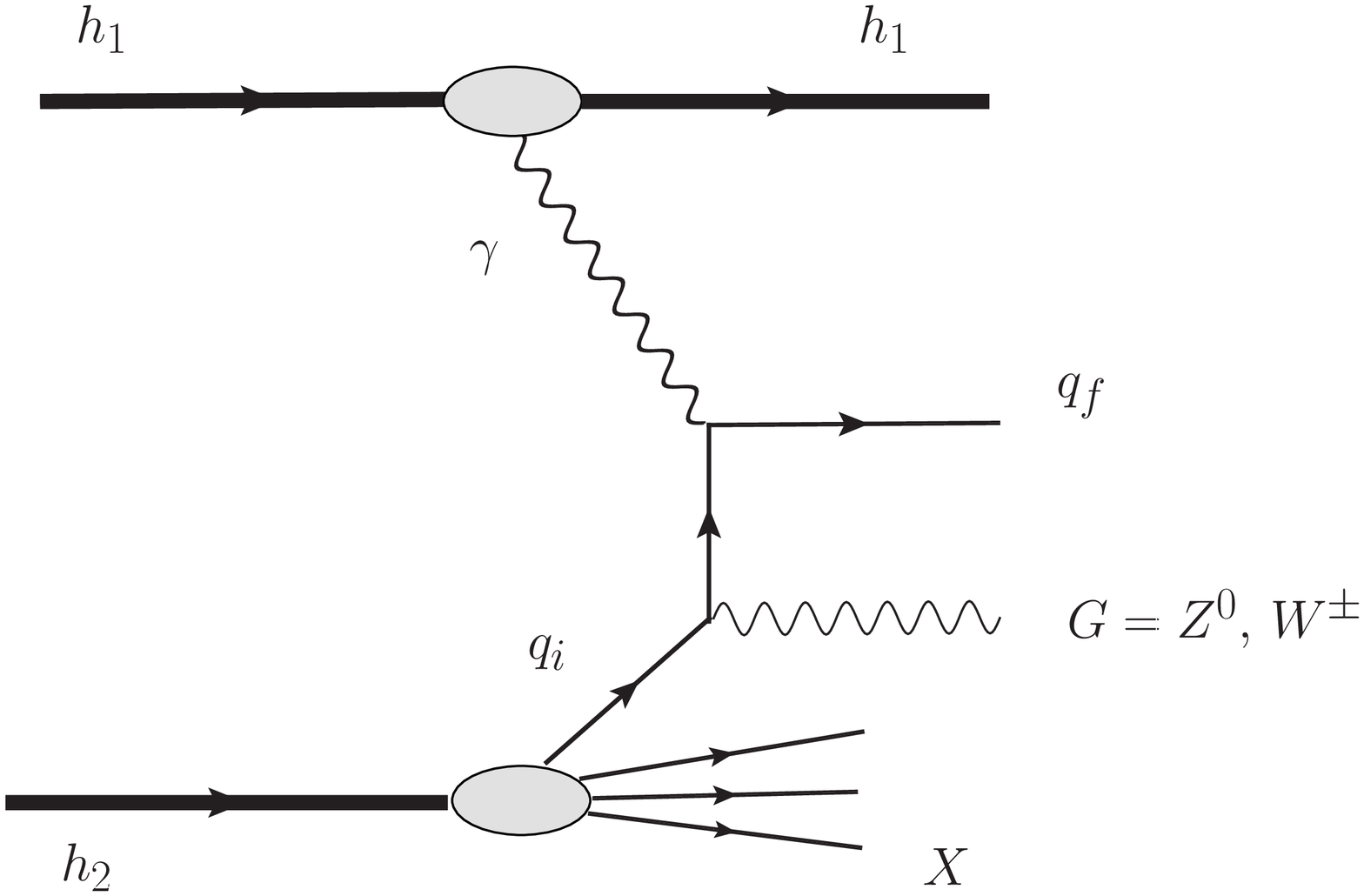,width=8.0cm}} &
{\psfig{figure=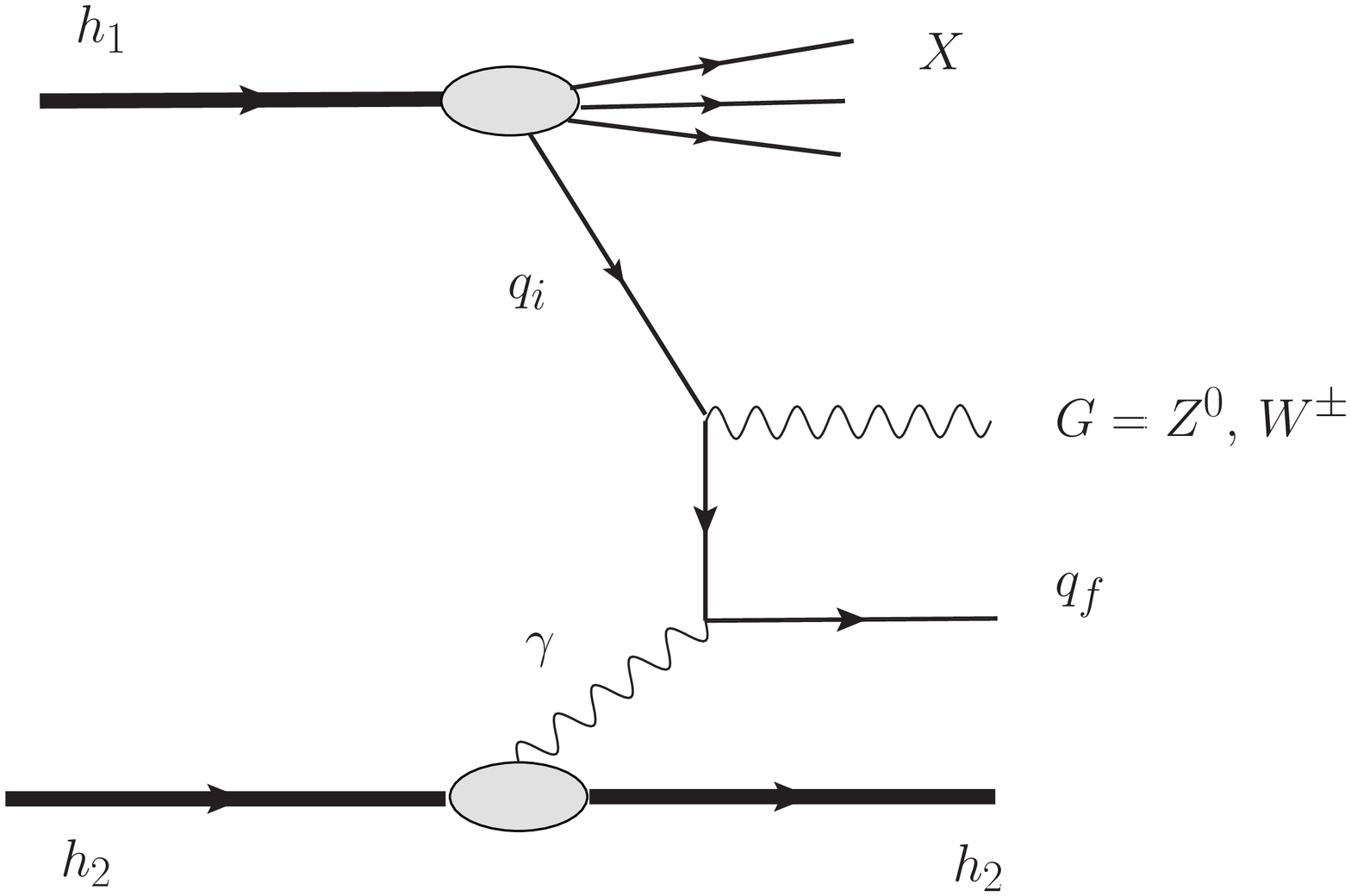,width=8.cm}}  
\end{tabular}                                                                                                                       
\caption{Photoproduction of massive gauge bosons in hadronic collisions.}
\label{fig:diagram}
\end{figure}

The   photoproduction of massive gauge bosons ($G = Z^0, \,W^{\pm}$) in hadronic  collisions is represented in Fig. \ref{fig:diagram}, where we have considered that both incident hadrons can be the source of photons. 
 The experimental signature for such events is the presence of one rapidity gap, associated to the photon exchange,  and one intact hadron in the final state. The associated total cross section is given by \cite{upc7}
\begin{equation}
   \sigma(h_1+h_2 \rightarrow h \otimes G + X) =  \int d \omega \,\,n_{h_1}(\omega)
   \, \sigma_{\gamma h_2 \rightarrow G X}\left(W_{\gamma h_2}  \right) + \int d \omega \,\, n_{h_2}(\omega)
   \, \sigma_{\gamma h_1 \rightarrow G X}\left(W_{\gamma h_1}  \right)\,  \; , 
\label{eq:sigma_pp}
\end{equation}
where $\otimes$ represents the presence of one rapidity gap in the final state, $\omega$ is the photon energy in the center-of-mass frame (c.m.s.), $n_{h_i}(\omega)$ is the equivalent photon flux for the hadron $h_i$, $W_{\gamma h}$ is the c.m.s. photon-hadron energy given by $W_{\gamma h}^2=2\,\omega\sqrt{s_{NN}}$, with
$\sqrt{s_{NN}}$ being  the c.m.s energy of the
hadron-hadron system. Considering the requirement that  photoproduction
is not accompanied by hadronic interactions (ultra-peripheral
collision) an analytic approximation for the equivalent photon flux of a nucleus can be calculated, which is given by \cite{upc7,upc1}
\begin{eqnarray}
n_{A}(\omega)= \frac{2\,Z^2\alpha_{em}}{\pi\,\omega}\, \left[\bar{\eta}\,K_0\,(\bar{\eta})\, K_1\,(\bar{\eta})- \frac{\bar{\eta}^2}{2}\,{\cal{U}}(\bar{\eta}) \right]\,
\label{fluxint}
\end{eqnarray}
where  $K_0(\eta)$ and  $K_1(\eta)$ are the
modified Bessel functions, $\bar{\eta}=\omega\,(R_{h_1}+R_{h_2})/\gamma_L$ and  ${\cal{U}}(\bar{\eta}) = K_1^2\,(\bar{\eta})-  K_0^2\,(\bar{\eta})$. Moreover, $\gamma_L$ is the Lorentz boost  of a single beam and we consider $R_p = 0.6$ fm and $R_A = 1.2 \, A^{1/3}$ fm in our calculations for $pPb$ and $PbPb$ collisions. For the photon spectrum associated to the proton we will assume that it  is given by  \cite{Dress},
\begin{eqnarray}
n_{p}(\omega) =  \frac{\alpha_{\mathrm{em}}}{2 \pi\, \omega} \left[ 1 + \left(1 -
\frac{2\,\omega}{\sqrt{s_{NN}}}\right)^2 \right] 
\left( \ln{\Omega} - \frac{11}{6} + \frac{3}{\Omega}  - \frac{3}{2 \,\Omega^2} + \frac{1}{3 \,\Omega^3} \right) \,,
\label{eq:photon_spectrum}
\end{eqnarray}
with the notation $\Omega = 1 + [\,(0.71 \,\mathrm{GeV}^2)/Q_{\mathrm{min}}^2\,]$ and $Q_{\mathrm{min}}^2= \omega^2/[\,\gamma_L^2 \,(1-2\,\omega /\sqrt{s_{NN}})\,] \approx (\omega/
\gamma_L)^2$. This expression  is derived considering the Weizs\"{a}cker-Williams method of virtual photons and using an elastic proton form factor (For more details see Refs. \cite{Dress,Kniehl}). 
 The photoproduction cross section for the $\gamma h \rightarrow G X$ process is given by \cite{WWg1,WWg2}
\begin{eqnarray}
\sigma_{\gamma h \rightarrow G X} (W_{\gamma h})=\int_{x_{min}}^{1}dx \sum_{q,\bar{q}}f_{q/h}(x,Q^{2})\,\hat{\sigma}_G(\hat{s}),
\end{eqnarray}
where $f_{q/h}$ are the parton distribution functions in the hadron target ($h = p$ or $Pb$), $x_{min}=M_G^2/W_{\gamma h}^2$,  $\hat{\sigma}_G$ is the cross section for the subprocess $\gamma q_i \rightarrow G q_f$ and $\hat{s}=x \cdot W_{\gamma h}^2$.
For $G = Z^0$ we have that $q_f = q_i$ (See Fig. \ref{fig:diagram}) and the subprocess cross section, $\hat{\sigma}_{Z}(\hat{s})$, is given at leading order by \cite{WWg1,WWg2,WWg3}
\begin{eqnarray}
\hat{\sigma}_Z & = &  \frac{\alpha G_{F}M_{Z}^{2}}{\sqrt{2}\,\hat{s}}\,
g_q^2e_q^2\, \left[ \left(1-2\hat{z}+2\hat{z}^{2}\right)\log \left(\frac{\hat{s}-M_{Z}^{2}}{\Lambda^{2}}\right) + \frac{1}{2}\left(1+2\hat{z}-3\hat{z}^{2}\right)\right],
\label{eq:sighz}
\end{eqnarray}
where $\hat{z}=M_{Z}^{2}/\hat{s}$, $g_q^{2}= \frac{1}{2}(1-4|e_q|x_W+8e_q^2x_W^2)$, $e_{q}$ is the quark charge and $x_W = 0.23$. 
On the other hand, for $G = W$, we have  $q_f \neq q_i$ and $\hat{\sigma}_W $ is given by \cite{WWg1,WWg2,WWg3}
\begin{eqnarray}
\hat{\sigma}_W  & = & \sigma_0 \, |V_{if}|^{2}\left\{(|e_{q}|-1)^{2}(1-2\hat{z}+2\hat{z}^{2})
\log({\hat{s}-M_{W}^{2}\over\Lambda^{2}}) 
 -  \left[(1-2\hat{z}+2\hat{z}^{2}) - 2|e_{q}|(2+2\hat{z}^{2}) - 1\right]\log{\hat{z}} \right.  \nonumber \\
& + &  \left. \left[ \frac{2}{\hat{z}} + \left(\frac{1}{2}
 +  {{3(1+|e_{q}|^{2})}\over{2}}\right)\hat{z}
 + 2|e_{q}| +  {|e_{q}|^{2}\over 2}\right](1-\hat{z}) \right\}
\label{eq:sighw}
\end{eqnarray}
where  $\sigma_0 ={{\alpha G_{F}M_{W}^{2}}\over{\sqrt{2}\hat{s}}}$, $\hat{z}=M_{W}^{2}/\hat{s}$ and quantities $V_{if}$ are the elements of the Cabibbo-Kobayashi-Maskawa (CKM) matrix.
The scale $\Lambda^{2}$ in the Eqs. (\ref{eq:sighz}) and (\ref{eq:sighw})  is a cutoff scale that regulates the singularity present at LO when the final state quark $q_f$ becomes collinear with the initial state photon.  Such  singularity does not occur at LO for a non-vanishing transverse momentum  $p_T$ of the massive gauge boson, since in this case the  $p_T$ of the boson has to be balanced by the final state quark. As in previous studies \cite{WWg2,WWg3,crismagno}, we will assume that $\Lambda = 0.4$ GeV.
Moreover, we will consider that the parton distribution functions for the proton are described by the CTEQ parametrization proposed in Ref. \cite{cteq}, while for photon -- nucleus interactions we will assume that the nuclear parton distributions can be expressed as $f_{q/A}(x,Q^{2}) = A \cdot R_q(x,Q^{2})   \cdot f_{q/p}(x,Q^{2})$, where the function  $R_q(x,Q^{2})$ parameterizes the nuclear effects in the parton distributions and is described by the EPPS16 parametrization \cite{epps}.

The rapidity distribution for the  photoproduction of massive gauge bosons in   $pp$, $pPb$ and $PbPb$ collisions can be calculated considering that the  rapidity $Y$ of the boson in the final state is directly related to the photon energy $\omega$ by the relation $Y\propto \ln \, ( \omega/m_{G})$.  Explicitly, the rapidity distribution is written down as,
\begin{eqnarray}
\frac{d\sigma}{dY} [h_1 + h_2 \rightarrow   h \otimes G + X] = \left[ \omega \, n_{h_1} (\omega) \, \sigma_{\gamma h_2 \rightarrow G\, X} \right]_{\omega_L} + 
\left[ \omega \,n_{h_2} (\omega) \, \sigma_{\gamma h_1 \rightarrow G\, X} \right]_{\omega_R}\,
\label{dsigdy}
\end{eqnarray}
where $\omega_L \, (\propto e^{Y})$ and $\omega_R \, (\propto e^{-Y})$ denote photons from the $h_1$ and $h_2$ hadrons, respectively. As the cross section increases with the energy, we have that the first term on the right-hand side of the Eq. (\ref{dsigdy}) peaks at positive rapidities while the second term peaks for negative rapidities. Consequently, given the photon flux, the study of the rapidity distribution can be used to constrain  the photoproduction cross section for a given energy. Moreover, the rapidity distributions for $pp$ and $PbPb$ collisions will be symmetric about midrapidity ($Y=0$). In contrast, for $pPb$ collisions, $d\sigma/dY$ will be asymmetric due to the dominance of the nuclear photon flux, which is proportional to $Z^2$.

\begin{figure}
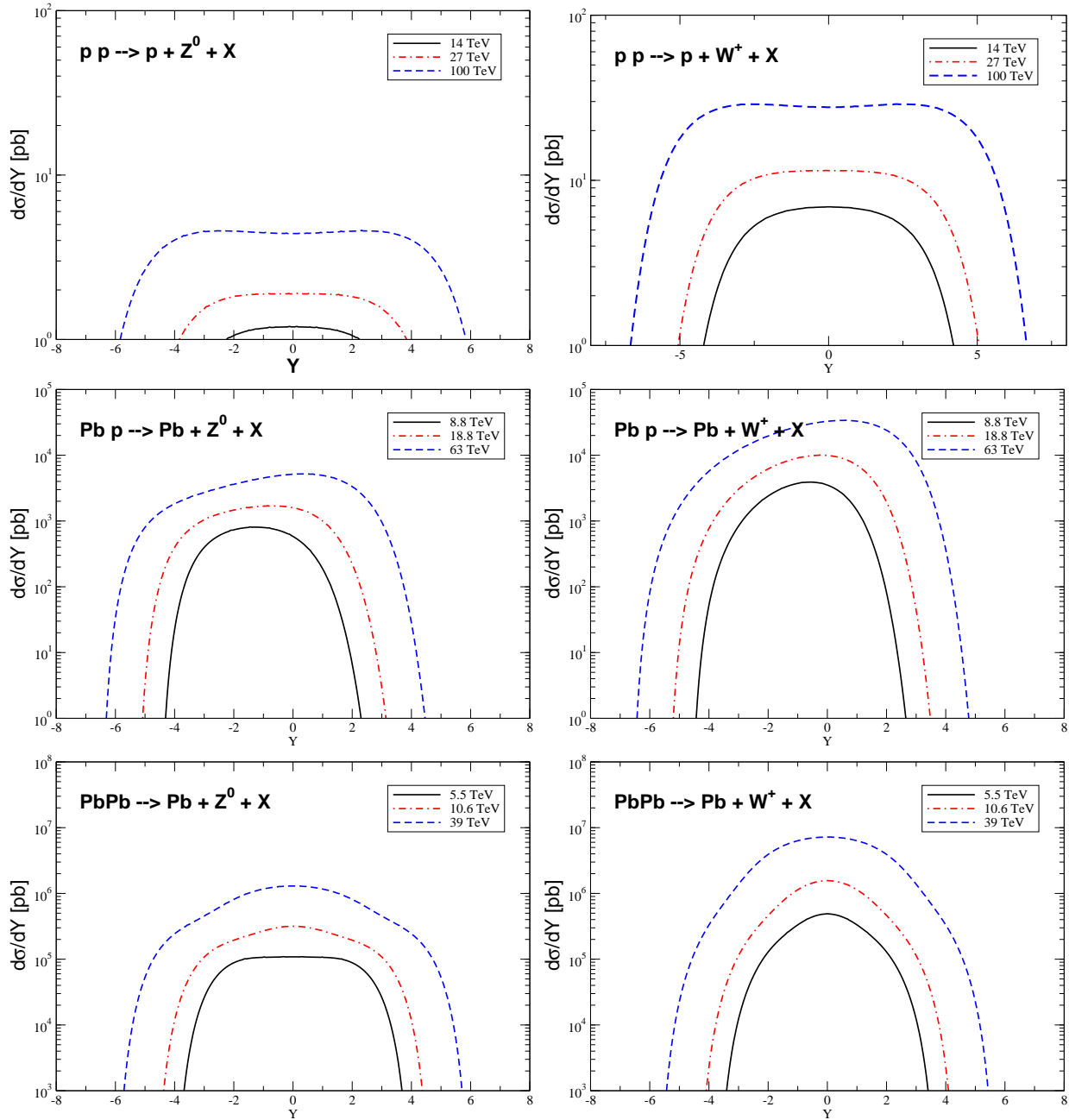

\begin{tabular}{cc}
\hspace{-1cm}
{\psfig{figure=Z0_pp.eps,width=8cm}} & 
{\psfig{figure=W_pp.eps,width=8cm}} \\ 
\hspace{-1cm}
{\psfig{figure=Z0_pPb.eps,width=8cm}}&
{\psfig{figure=W_pPb.eps,width=8cm}} \\
\hspace{-1cm}
{\psfig{figure=Z0_PbPb.eps,width=8cm}}&
{\psfig{figure=W_PbPb.eps,width=8cm}} \\
\end{tabular}                                                                                                                       
\caption{Rapidity distributions for the  $Z^0$ (left panels) and $W^+$ (right panels) photoproduction in $pp$ (upper panels), $pPb$ (central panels) and $PbPb$ (lower panels) collisions.}
\label{fig:rapdist}
\end{figure}

In Fig. \ref{fig:rapdist} we present our predictions for the rapidity distributions 
for the  $Z^0$ (left panels) and $W^+$ (right panels) photoproduction in $pp$ (upper panels), $pPb$ (central panels) and $PbPb$ (lower panels) collisions. We consider the planned center -- of -- mass energies for the next run of the LHC, as well for the future High -- Energy LHC  \cite{he_lhc} and Future Circular Collider \cite{fcc}. We have that the predictions for the $W^+$ production are larger than for the $Z^0$ one, which is expected from the results for the photoproduction of massive gauge bosons presented in Refs. \cite{WWg2,crismagno}. Moreover, we have that for larger energies the  rapidity distribution increases  and becomes wider in rapidity. As expected from our previous discussion, the distribution is symmetric for $pp$ and $PbPb$ collisions and asymmetric for $pPb$ one, with the maximum occurring for positive rapidities. Finally, the predictions for midrapidities ($Y \approx 0$) increase with the energy and are a factor $ \approx 10^3$ ($10^6$) larger in $pPb$ ($PbPb$) collisions than in $pp$ collisions.

\begin{center}
\begin{table}
\begin{tabular}{|c|c|c|c|c|}
\hline 
{\bf $pp$ collisions} & $\sigma (Z^{0}) [pb]$ & $\#$ events ($Z^0 \rightarrow \mu^+ \mu^-$)  & $\sigma (W^{+}) [pb]$& $\#$ events ($W^+ \rightarrow \mu^+ \nu_{\mu}$) \tabularnewline
\hline 
$\sqrt{s}$ = 14 TeV & 5.67  & 190.0  & 32.35 &   3438.0   \tabularnewline
\hline 
$\sqrt{s}$ = 27 TeV & 9.36 & 315.0 & 56.30 &  5984.0    \tabularnewline
\hline 
$\sqrt{s}$ = 100 TeV & 22.69 & 764.0 & 142.80 & 15179.0  \tabularnewline
\hline
\hline 
\hline 
{\bf $Pbp$ collisions} & $\sigma (Z^{0}) [pb]$ & $\#$ events ($Z^0 \rightarrow \mu^+ \mu^-$)  & $\sigma (W^{+}) [pb]$ & $\#$ events ($W^+ \rightarrow \mu^+ \nu_{\mu}$) \tabularnewline
\hline 
$\sqrt{s}$ = 8.8 TeV & 2.29 $\times 10^3$  &  77.0  &  11.17 $\times 10^3$ & 1187.0     \tabularnewline
\hline 
$\sqrt{s}$ = 18.8 TeV & 5.99 $\times 10^3$ & 201.0  &  34.16 $\times 10^3$ &  3631.0  \tabularnewline
\hline 
$\sqrt{s}$ = 63 TeV & 21.44 $\times 10^3$ & 20940.0  &  135.88 $\times 10^3$ & 418877.0 \tabularnewline
\hline
\hline
\hline 
{\bf $PbPb$ collisions} & $\sigma (Z^{0}) [pb]$ & $\#$ events ($Z^0 \rightarrow \mu^+ \mu^-$) & $\sigma (W^{+}) [pb]$ & $\#$ events ($W^+ \rightarrow \mu^+ \nu_{\mu}$) \tabularnewline
\hline 
$\sqrt{s}$ = 5.5 TeV & $0.49\times 10^6$  & 165.0 & $1.40\times 10^6$ & 1488.0    \tabularnewline
\hline 
$\sqrt{s}$ = 10.6 TeV & $1.24\times 10^6$ & 417.0 & $4.74\times 10^6$ & 5038.0    \tabularnewline
\hline 
$\sqrt{s}$ = 39 TeV & $5.26\times 10^6$ & 19487.0  & $27.60\times 10^6$ & 322726.0 \tabularnewline
\hline
\end{tabular}
\caption{Cross sections  and associated number of events in the leptonic decay mode for the photoproduction of massive gauge bosons in $pp/pPb/PbPb$ collisions considering the typical rapidity range covered by a central detector ($|Y| \le 2$).}
\label{tab:central}
\end{table}
\end{center}

\begin{center}
\begin{table}
\begin{tabular}{|c|c|c|c|c|}
\hline 
{\bf $pp$ collisions} & $\sigma (Z^{0})[pb]$ & $\#$ events ($Z^0 \rightarrow \mu^+ \mu^-$)& $\sigma (W^{+}) [pb]$ & $\#$ events ($W^+ \rightarrow \mu^+ \nu_{\mu}$)\tabularnewline
\hline 
$\sqrt{s}$ = 14 TeV & 1.47 & 49.0 & 8.69 &   923.0   \tabularnewline
\hline 
$\sqrt{s}$ = 27 TeV & 3.28 & 110.0  & 20.44 &  2172.0   \tabularnewline
\hline 
$\sqrt{s}$ = 100 TeV & 10.95 & 368.0  & 69.35 & 7371.0  \tabularnewline
\hline
\hline 
\hline 
{\bf $Pbp$ collisions} & $\sigma (Z^{0}) [pb]$ & $\#$ events ($Z^0 \rightarrow \mu^+ \mu^-$) & $\sigma (W^{+}) [pb]$ & $\#$ events ($W^+ \rightarrow \mu^+ \nu_{\mu}$) \tabularnewline
\hline 
$\sqrt{s}$ = 8.8 TeV & 0.0012 $\times 10^3$& 0.04  & 0.015 $\times 10^3$ & 2.0      \tabularnewline
\hline 
$\sqrt{s}$ = 18.8 TeV & 0.070 $\times 10^3$  & 2.35 & 0.67 $\times 10^3$ &  71.0   \tabularnewline
\hline 
$\sqrt{s}$ = 63 TeV & 2.56 $\times 10^3$ & 2500.0 & 20.18 $\times 10^3$ & 62208.0 \tabularnewline
\hline
\hline
\hline 
{\bf $PbPb$ collisions} & $\sigma (Z^{0}) [pb]$ & $\#$ events ($Z^0 \rightarrow \mu^+ \mu^-$) &  $\sigma (W^{+}) [pb]$ &  $\#$ events ($W^+ \rightarrow \mu^+ \nu_{\mu}$)\tabularnewline
\hline 
$\sqrt{s}$ = 5.5 TeV & $0.068 \times 10^6$ & 22.0  & $0.067\times 10^6$ &  71.0    \tabularnewline
\hline 
$\sqrt{s}$ = 10.6 TeV & $0.21\times 10^6$ & 70.0 & $0.32\times 10^6$ &  340.0    \tabularnewline
\hline 
$\sqrt{s}$ = 39 TeV & $1.06\times 10^6$ & 3927.0 & $3.43\times 10^6$ & 40106.0 \tabularnewline
\hline
\end{tabular}
\caption{Cross sections and associated number of events in the leptonic decay mode for the photoproduction of massive gauge bosons in $pp/pPb/PbPb$ collisions considering the typical rapidity range covered by a forward detector ($2 \le Y \le 4.5$).}
\label{tab:forward}
\end{table}
\end{center}

In the Tables \ref{tab:central} and \ref{tab:forward} we present our predictions for the total cross sections and number of events considering the rapidity range covered by a typical central detector ($|Y| \le 2.0$), as the ATLAS and CMS detectors, as well as for a forward detector  ($2.0 \le Y \le 4.5$) as the LHCb detector.
For $pp/pPb/PbPb$ collisions we predict cross sections of the order of $pb/nb/\mu b$, with the results for a forward detector being, in general, smaller by a factor $\ge 3$ than for a central detector. In order to estimate the number the events we will consider that the expected integrated luminosities in $pp/pPb/PbPb$ collisions for the next run of LHC and HE -- LHC are ${\cal{L}} = 1\,fb^{-1}/\,1\,pb^{-1}/\,10\,nb^{-1}$, while for the FCC are ${\cal{L}} = 1\,fb^{-1}/\,29\,pb^{-1}/\,110\,nb^{-1}$. Moreover, we will also take into account the leptonic decay mode, with the associated branching ratios for the processes $W^+ \rightarrow \mu \nu_{\mu}$ and $Z^0 \rightarrow \mu^+ \mu^-$ being   10.63 \%     and 3.3658 \%       , respectively. The results presented in  
the Tables \ref{tab:central} and \ref{tab:forward} indicate that the number of events is 
 large, especially for $W^+$ production at midrapidities and FCC energies. In principle, such large number will allow to perform the search of tiny effects, as those associated to the presence of anomalous couplings between the gauge bosons, which are a signal of BSM physics (See e.g. Refs. \cite{crismagno,Dubinin,WWg3}). Such aspect will be investigated in a future publication.
 

 Two comments are in order. First, in our analysis the contribution for the massive gauge boson production associated to the hadronic component of the photon, denoted resolved contribution, was not included. Such contribution implies that  a $G + jet$ final state can be produced  via e.g. the $q \bar{q} \rightarrow g G$ subprocess. Previous studies have demonstrated that the resolved contribution slightly increases the magnitude of the cross section and  is negligible for the production of a massive gauge boson with a large transverse momentum \cite{WWg2}. Second, our estimates were obtained at leading order. The next --  to -- leading order (NLO) corrections are predicted  to  increase the LO direct cross section by $\approx 10 \%$ \cite{Diener:2002if}. As both corrections increase the cross sections, the results presented in this letter can be considered a lower bound for the magnitude of the number of events expected at the LHC, HE -- LHC and FCC. We plan to perform a more detailed analysis, taking into account of these corrections, in a forthcoming study.

As a summary, in this letter we have performed an exploratory study and estimated, for the first time, the $Z^0$ and $W^+$ photoproduction in hadronic collisions at the LHC, HE -- LHC and FCC energies. Our study is strongly motivated by the high photon -- hadron luminosity present in hadronic collisions at high energies, which  become feasible  the experimental analysis of different final state  that can be used to test some of the more important properties of Standard Model (SM) as well to search by BSM physics.  We have estimated the rapidity distributions for $pp$, $pPb$ and $PbPb$ collisions, cross sections for the rapidity ranges covered by central and forward detectors, as well the corresponding number of events in the leptonic decay mode. Our results indicate that the number of events is large enough to allow a future experimental analysis to probe the Standard Model predictions and possible scenarios of the BSM physics.

\begin{acknowledgments}
VPG thank the members of the Faculty of Nuclear Sciences and Physical Engineering of the Czech Technical University in Prague by the warm hospitality during the completion of this work.
This work was  partially financed by the Brazilian funding
agencies CNPq,   FAPERGS and INCT-FNA (process number 
464898/2014-5).
\end{acknowledgments}

 \end{document}